\documentclass[12pt,preprint2]{aastex}

\shorttitle{Binarity and Dust Processing}
\shortauthors{Pascucci et al.}

\begin{document}


\title{Medium--separation binaries do not affect the first steps of planet formation}



\author{I. Pascucci, D. Apai, E. E. Hardegree--Ullman, J. S. Kim, M. R. Meyer,}
\affil{Steward Observatory, The University of Arizona, Tucson, AZ 85721}
\and
\author{J. Bouwman}
\affil{Max Planck Institute for Astronomy, K\"{o}nigstuhl 17, D-69117, Heidelberg, Germany}




\begin{abstract}
The first steps of planet formation are marked by the growth and 
crystallization of sub--micrometer--sized dust grains accompanied 
by dust settling toward the disk midplane.
In this paper we explore whether the first steps of planet formation are affected by the presence of
medium--separation stellar companions. We selected two large samples of disks around single and binary T~Tauri stars in Taurus that are thought to have only a modest age spread of a few Myr. The companions of our binary sample are at projected separations between $\sim$10 and 450\,AU with masses down to about 0.1\,M$_{\sun}$. We used the strength and shape of the 10\,\micron{} silicate emission feature as a proxy for grain growth and for crystallization respectively. The degree of dust settling was evaluated from the ratio of fluxes at two different mid--infrared wavelengths. 
We find no statistically significant difference between the distribution of 10\,\micron{} silicate emission features from single and binary systems. In addition,  the distribution of disk flaring is indistinguishable between the single and binary system samples. These results show that  the first steps of planet formation are not affected by the presence of a companion at tens of AU.

\end{abstract}


\section{Introduction}
Two--third of the G stars in the  solar neighborhood are members of multiple--star systems
(e.g. \citealt{dm91}). These binaries and multiple systems  are often found to harbor giant planets (e.g. \citealt{bondes07}). Similarly, young low--mass pre--main sequence stars are very frequently members of multiple systems, 
mostly binaries \citep{mathieu00,duchene07}. This suggests that planet formation around single stars such as our Sun may be atypical and urges us to understand the effects of stellar companions on planet formation. We  tackle this question from an observational point of view. Because numerical simulations of grain agglomeration suggest short timescales for the formation of planetesimals (only a few 10$^4$\,yr, e.g. \citealt{beckwith00}), it is crucial to know how  stellar companion(s) affect the dust processing in the first few million years. 

Grain growth and the settling of dust grains towards the disk midplane are thought to represent the first steps in the planet--formation process (e.g. \citealt{ls93} for a review). The study of disks around intermediate--mass stars 
also indicates a link between grain growth and crystallinity. High crystallinity was found 
only when grains larger than the dominant sub--micron interstellar grains were present \citep{van05}. In the context of these findings and dust evolution models (e.g. \citealt{dd04}), older disks are expected to have more processed dust  
(larger grains and crystals) than younger disks. In addition, their disk structure should be flatter because of the gradual settling of 
large dust grains towards the disk midplane.
However, recent observations show that the degree of dust processing can be very different 
even for coeval disks around stars of similar spectral type in the same star--forming region 
(e.g. \citealt{prz03,apai05}). This demonstrates that dust evolution is 
not uniquely controlled by stellar age and luminosity but at least one additional parameter 
is present. 

There are two studies suggesting that stellar multiplicity could play a major role in the initial dust processing.
\citet{meeus03} found that among three coeval T~Tauri disks in the Chamaeleon~I 
star--forming region the closest binary system (projected separation of $\sim$120\,AU) sports the strongest contribution from large 
($\sim$\,2\,\micron) grains and has the highest crystalline mass fraction. Similarly,  \citet{sterzik04} 
pointed out that the disk of a young brown dwarf with a companion at $\ga$\,30\,AU shows  more processed dust than the disk around a single brown dwarf. 
Although the small samples inhibit any firm conclusions, these results suggest that companions might trigger rapid dust evolution.
Intuitively this may happen in different ways. A companion could speed up dust 
evolution by dynamically stirring the circumstellar dust grains and leading to an enhanced 
collision and grain growth rate (e.g. \citealt{dubrulle95}).  In addition, the dynamical stirring may lead to an increased  mixing that could also expose larger amounts of dust to temperatures high enough ($\ge$\,800\,K) to be crystallized. 

In this paper we compare two carefully constructed samples of  disks 
around single and binary\footnote{a few of the systems in our study are triple or quadruple systems. For simplicity, we refer to the whole sample as binaries)} stars with a narrow age spread to test the hypothesis that binary systems have disks with more
processed dust and flatter structures.  
In Sect.~\ref{S:sample} we describe our samples. The data reduction of 
the {\it Spitzer} spectra and of the 24\,\micron{} MIPS photometry is presented in Sect.~\ref{S:redu}.
We summarize our results in Sect.~\ref{S:results} and discuss in Sect.~\ref{S:discussion} their implications on planet formation in single and binary systems.




\section{Sample Definition}\label{S:sample}
Testing whether stellar companions promote the first steps of planet formation requires: a) two 
coeval samples of disks around single and binary stars with precisely determined 
multiplicity; b) identical spectral type distributions for the two samples; c) objects in regions with no diffuse PAH emission that could contaminate the dust emission 
features of the targets; d) disks with faint or no PAH emission features (for the same reason as b). 
The Taurus--Auriga star--forming region has a reasonably complete census of its pre--main--sequence 
stellar population and meets best our requirements among the nearby star--forming regions. 
To satisfy criteria d) we selected disks around low--mass stars because they have an order of magnitude lower PAH emission than disks around intermediate--mass stars  \citep{geers06}. 
The  observed age distribution of low--mass stars in Taurus can be well approximated with a gaussian centered at 1.6\,Myr and a spread no larger than about 2--3\,Myr \citep{hartmann01}. This narrow age spread minimizes any possible trend of disk evolution with age.
In summary our sample of disks around single and binary stars is drawn from the Taurus--Auriga population of low--mass T Tauri stars (TTSs) with circumstellar disks and ages between $\sim$1--3\,Myr.

Because it is critically important to include only stars with known multiplicity, 
we first selected each target based on a combination of available 
high--resolution imaging and interferometry, spectroscopy, and radial velocity measurements 
\citep{ghez93,leinert93,simon95,wg01}. In this context, we refer to a star as {\it single} if it has no known companions with brightness above the detection limit within 10\arcsec{} (or $\sim$\,1400\,AU at the $\sim$\,140\,pc distance  of the Taurus--Auriga star--forming region). The typical  detection limit of high--resolution imaging surveys is 2--3\,mag fainter than the primary star in K--band, which corresponds to a very--low mass star of $\sim$0.1\,M$_{\sun}$ at $\sim$2\,Myr according to the Baraffe et al. 1998 isochrones. Only a few binary star systems have been observed with techniques reaching higher contrast (see, e.g. the discovery of a brown dwarf companion to DH~Tau by \citealt{itoh05}).
The typical smallest separation resolvable with the imaging surveys is $\sim$\,0\farcs1 ($\sim$10\,AU), meaning that a few unresolved close binaries may have been classified as single.
Of these, binaries with periods $<$\,100\,days can only modestly contaminate our single star sample since their frequency in pre--main sequence stars is 8\%$\pm$3\% \citep{mathieu00}. Due to the lack of general understanding of the separation distribution of close pre--main sequence binaries, it is not possible to determine the exact level of contamination. However, the required infrared excess (see below) ensures circumstellar dust within a few AU, likely inconsistent with the presence of a close stellar companion.
We also excluded all known spectroscopic binaries in Taurus \citep{mathieu94,jm97} both from the sample of single and binary stars.
Our binary sample consists of stars  with companions between $\sim$0\farcs1 and 3\,\arcsec{} thus covering the population of medium--separation stellar companions at projected separations between $\sim$14--420\,AU.
After defining the samples, we searched for available near-- and mid--infrared data 
\citep{kh95,stassun01,hartmann05,mccabe06,furlan06} to select only those targets that have excess 
emission indicative of a circumstellar disk. These selection criteria resulted in a sample of mostly classical TTSs for the single and for the primary star in the multiple systems.
Only  CZ~Tau and IQ ~Tau in our sample are classified as weak--lined TTSs\footnote{based on their narrow H$\alpha$ emission lines \citep{hb88}} 
\citep{wg01,mccabe06}. The T Tauri types for the components of the binary sample are mostly classical with the exception of FX~TauB, IS~TauB, V710~TauB, and V807~TauB that are weak--lined TTSs \citep{duchene99,wg01, harken03}.
Finally, we selected only systems with fluxes $\gtrsim$\,0.1\,Jy to ensure high  signal--to--noise around the 10\,\micron{} silicate emission feature.

\begin{figure*}
\includegraphics[angle=90,scale=0.7]{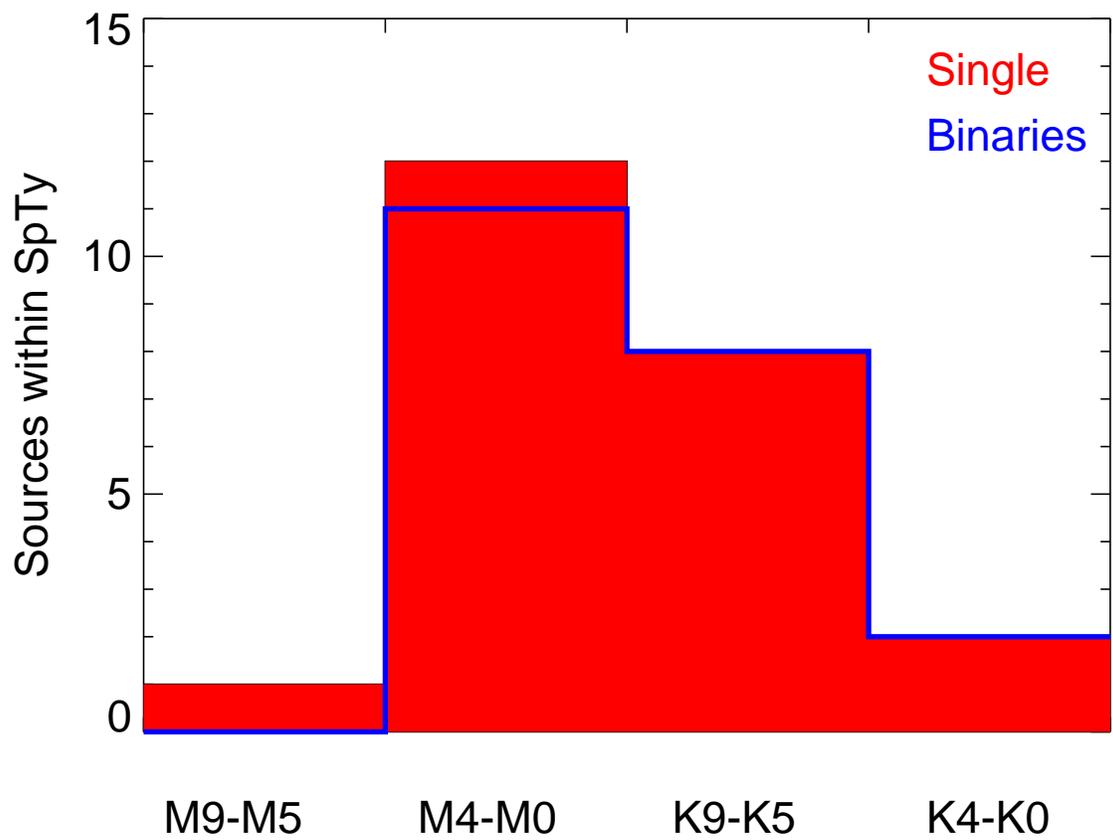}
\caption{Histogram of the spectral types for the single and binary systems. The spectral type distributions of the single stars and of the primary stars in the binary sample are almost identical.  \label{F:spty}}
\end{figure*}

Our list contains 44 sources, of which 23 are disks around single stars and 21 are multiple systems  (see Tables~\ref{T:single} and \ref{T:binary}). 
Spectral types for  the stellar companion(s) are available for 18 out of 21 objects. Most companions have M spectral type (see Table~\ref{T:binary}).
The single stars (or the primaries of multiple systems) are mostly between K5 
and M4 spectral type. Note also that the distribution of spectral types for the single stars is almost identical to that of the primary component in multiple systems (Fig.~\ref{F:spty}). In addition, the fairly narrow range in spectral type minimizes any difference in the dust 
composition due to luminosity effects. In fact the recent study by \citet{kessler06} shows that the shape and strength of the 10\,\micron{} silicate features of K versus M stars are consistent with being drawn from the same population.
Most of our multiple systems are binaries with separations less than 1\arcsec{} (see Table~\ref{T:binary}). Seventeen of them have circumprimary and circumsecondary disks based on the K--L, K--N colors or resolved IRAC photometry of both components 
\citep{wg01,hartmann05,mccabe06}. DF~Tau~B and FO~Tau~B are the closest binary components in our sample at $\sim$\,0\farcs1. They are also very likely surrounded by circumstellar disks based on accretion signatures like the H$\alpha$ equivalent width and the amount of optical veiling \citep{harken03}. Finally, V710~Tau~B and V807~Tau~B have K--N and K--L colors consistent with photospheric emission suggesting that they do not have circumstellar disks  \citep{wg01,mccabe06}.  A summary of the disk configurations is provided in column seven of Table~\ref{T:binary}. 
Resolved mid--infrared photometry is available for 11 out of 21 binary sources \citep{mccabe06}. Another 8 sources have resolved L--band photometry \citep{wg01}, while the two closest binaries FO Tau and DF Tau have been resolved only in K--band \citep{wg01}. For at least two--third of the binary sample the {\it Spitzer}/IRS spectrum is dominated by the disk around the primary star as indicated by the near-- and mid--infrared  flux ratios of the primary and secondary components (column eight of Table~\ref{T:binary}). We also note that our selection criteria did not exclude possibly more evolved disks such as transition disks. GM~Aur and DM~Tau in our list of single stars are among the best studied transition disks (see, e.g. \citealt{calvet05}).

\begin{deluxetable}{cl c cc}
\tabletypesize{\scriptsize}
\tablecaption{Main properties of the selected single TTSs.  \label{T:single}}
\tablewidth{0pt}
\tablehead{
\colhead{\#}&\colhead{Source}&\colhead{2MASS~J}&\colhead{Adopted} & \colhead{Ref} \\
\colhead{} & \colhead{} & \colhead{} &  \colhead{SpT} & \colhead{(SpT)}}
\startdata
	S1 & AA~Tau & 04345542+2428531  &  K7 & 1, 2 \\
	S2 & BP~Tau & 04191583+2906269	 &  K7  & 2, 3\\
	S3 & CI~Tau  & 04335200+2250301	 &  K7 & 1, 2 \\
	S4 & CW~Tau& 04141700+2810578  &  K3 & 1, 2 \\
	S5 & CX~Tau & 04144786+2648110	 &  M2.5 &1, 2 \\
	S6 & CY~Tau & 04173372+2820468	 &  M2 & 2, 3 \\
	S7 & DE~Tau & 04215563+2755060	 &  M1 & 2, 3\\
	S8 & DH~Tau\tablenotemark{\ast} & 04294155+2632582	 &  M2 & 2, 4\\
	S9 & DL~Tau & 04333906+2520382	 &  K7 & 1, 2 \\
	S10 & DM~Tau\tablenotemark{\blacklozenge} & 04334871+1810099& M1 & 1, 2 \\
	S11 & DN~Tau & 04352737+2414589 & M0 & 1, 2\\
	S12 & DO~Tau & 04382858+2610494 & M0 & 1, 2\\
	S13 & DP~Tau & 04423769+2515374 & M0.5 &1, 2\\
	S14 & DS~Tau & 04474859+2925112 & K5 &1, 2\\
	S15 & FM~Tau  & 04141358+2812492 & M1 & 2, 4\\
	S16 & FN~Tau & 04141458+2827580 & M5 & 2\\
	S17 & FZ~Tau\tablenotemark{\ast} & 04323176+2420029 & M0 & 2, 4, 5\\
	S18 & GI~Tau\tablenotemark{\ast} & 04333405+2421170	 &  K7 & 2, 4\\
	S19 & GK~Tau\tablenotemark{\ast} & 04333456+2421058&  K7 & 2, 4 \\
	S20 & GM~Aur\tablenotemark{\blacklozenge} & 04551098+3021595&  K3 & 1, 2\\
	S21 & IP~Tau & 04245708+2711565	&  M0 & 1, 2\\
	S22 & IQ~Tau & 04295156+2606448& M0.5 & 1, 2\\
	S23 & LkCa~15 & 04391779+2221034& K5 & 1, 2\\
\enddata
\tablenotetext{\ast}{These stars have companions at projected separations just above 10\arcsec{} \citep{hartigan94} :
DH~Tau/DI~Tau (15\arcsec); FY~Tau/FZ~Tau (16\farcs9);
GK~Tau/GI~Tau (12\farcs9)}
\tablenotetext{\blacklozenge}{Known transition disks (see, e.g. \citealt{calvet05}).}
\tablecomments{The 2MASS source name includes the J2000 sexagesimal, equatorial position in the form: hhmmssss+ddmmsss \citep{cutri03}.}
\tablerefs{
(1) \citealt{hb88}; (2) \citealt{kh95}; (3) \citealt{ss94};
(4) \citealt{hartigan94}; (5) \citealt{wg01}
}
\end{deluxetable}

\begin{deluxetable}{cl c cccccc}
\tabletypesize{\scriptsize}
\tablecaption{Main properties of the selected binary TTSs. \label{T:binary}}
\tablewidth{0pt}
\tablehead{
\colhead{\#}&\colhead{Source}&\colhead{2MASS~J}&\colhead{Adopted} & \colhead{Separation} & \colhead{Ref}  & \colhead{Disk} & \colhead{Flux ratio}& \colhead{Ref}\\
\colhead{} & \colhead{} & \colhead{} &  \colhead{SpT\tablenotemark{\clubsuit}} & \colhead{(arcsec)} & \colhead{(SpT)} &
\colhead{configuration\tablenotemark{\spadesuit}} & \colhead{{\tiny(filter)}\tablenotemark{\blacklozenge}}&\colhead{(flux ratio)}}
\startdata
	B1 & CoKu~Tau3 A-B & 04354093+2411087 & M1 & 2.05 & 1 & cp+cs & 4.7 {\tiny(L)} & 3\\
	B2 & CZ~Tau A-B       & 04183158+2816585 & M1.5 & 0.32 & 1 & cp+cs & 1.71 {\tiny(L)} & 10\\
	B3 & DD~Tau A-B       & 04183112+2816290 & M3+M3 & 0.56 & 2, 3 & cp+cs &1.75 {\tiny(N)}& 10\\
	B4 & DF~Tau A -B       & 04270280+2542223 & M0.5+M3& 0.09& 4, 3  & cp+cs & 1.6 {\tiny(K)}& 3\\
	B5 & DK~Tau A-B       & 04304425+2601244 & K9+M1 & 2.30 & 1, 5  & cp+cs& 8.53 {\tiny(SiC)}& 10\\	
	B6 & FO~Tau A-B        & 04144928+2812305 & M2+M2 & 0.15 & 4, 3  & cp+cs& 1.7 {\tiny(L)}& 3\\	
	B7 & FS~Tau A-B       & 04220217+2657304 & M1+M4 & 0.23 & 4, 3  & cp+cs& 5 {\tiny(L)}& 3\\
	B8 & FV~Tau A-B       & 04265352+2606543 & K5+K6 & 0.72 & 4, 2, 3  & cp+cs&2.2 {\tiny(N)}& 10\\
	B9 & FX~Tau A-B       & 04302961+2426450 & M1+M4 & 0.89 & 1, 5  & cp+cs&2.8 {\tiny(N)}& 10\\
	B10 & GG~Tau Aa-Ab& 04323034+1731406 & K7+M0.5 & 0.25 & 6 & cp+cs&1.03 {\tiny(N)}& 10\\
	B11 & GH~Tau A-B     & 04330622+2409339 & M1.5+M2 & 0.31 & 3 & cp+cs&1.45 {\tiny(N)}& 10\\
	B12 & GN~Tau A-B     & 04392090+2545021 & M2.5 & 0.33 & 7  & cp+cs&1.53 {\tiny(L)} & 10\\
	B13 & HK~Tau A-B     & 04315056+2424180 & M1+M2 &2.34 & 1, 5 & cp+cs&30 {\tiny(SiC)}& 10\\
	B14 & HN~Tau A-B     & 04333935+1751523 & K5+M4 & 3.11 & 2  & cp+cs&65 {\tiny(L)}& 3\\
	B15 & IS~Tau A-B       & 04333678+2609492 & K7+M4.5 & 0.22 & 8, 3  & cp+cs&9 {\tiny(L)}& 3\\
	B16 & IT~Tau A-B       & 04335470+2613275 & K3+M4 & 2.39 & 5  & cp+cs&2.95 {\tiny(SiC)}& 10\\
	B17 & RW~Aur A-B    & 05074953+3024050 & K1+K5 & 1.42 & 9, 5  & cp+cs&13.63 {\tiny(N)}& 10\\
	B18 & UY~Aur A-B     & 04514737+3047134 & K7+M2 & 0.88 & 1, 5 & cp+cs&2.07 {\tiny(N)}& 10\\
	B19 & V710~Tau A-B & 04315779+1821380 & M0.5+M2 & 3.17 & 2 & cp&12.7 {\tiny(SiC)}& 10\\
	B20 & V807~Tau A-B & 04330664+2409549 & K7+M3 & 0.3 & 4, 2, 3 & cp&3.6 {\tiny(L)}& 3 \\
	B21 & V955~Tau A-B & 04420777+2523118 & K5+M1 & 0.33  & 3  & cp+cs&5.6 {\tiny(L)}& 3\\
\enddata
\tablenotetext{\clubsuit}{The second spectral type, when available, is for the secondary star.}
\tablenotetext{\spadesuit}{'cp' stands for circumprimary disk while 'cs' stands for circumsecondary disk. These disk configurations have been determined via resolved optical and infrared photometry, see Sect.~\ref{S:sample} for details.}
\tablenotetext{\blacklozenge}{Flux ratios are calculated as primary/secondary. In parenthesis we provide the filter at which the flux ratio has been calculated (the SiC filter is centered at 11.8\,\micron{}, \citealt{mccabe06}). }
\tablecomments{The 2MASS source name includes the J2000 sexagesimal, equatorial position in the form: hhmmssss+ddmmsss \citep{cutri03}.}
\tablerefs{
(1) \citealt{leinert93}; (2) \citealt{hartigan94}; (3) \citealt{wg01};
(4) \citealt{cohen79}; (5) \citealt{duchene99};
(6) \citealt{white99};  (7) \citealt{wb03};
(8) \citealt{martin94};  (9) \citealt{mundt82}; (10) \citealt{mccabe06}
}
\end{deluxetable}

\section{Data Reduction}\label{S:redu}
We use the low--resolution {\it Spitzer} spectra around the 10\,\micron{} silicate emission feature to characterize the degree of dust processing in  single and binary systems. Observations at wavelengths longer than $\sim$\,20\,\micron{} are necessary to
trace the evolution of the disk structure. For this, we prefer to use the MIPS 24\,\micron{} photometry that is available for all (except 6) targets rather than the spectra acquired with the long--low module of the IRS, which is lacking for 14 targets.
In the following we describe the data reduction of the {\it Spitzer} IRS spectra and of the 24\,\micron{} MIPS photometric data.

\subsection{{\it Spitzer} IRS low--resolution spectra}
The IRS data presented in this paper have been acquired as part of the IRS/GTO program \citep{furlan06}
and became available to the community at the end of the year 2005. 
Only six of the objects we selected (CY~Tau, DS~Tau, FM~Tau, IS~Tau, V710~Tau, and V807~Tau)
were observed in staring mode, with the targets placed in two nod positions along the spatial direction of the slit at 1/3 and 2/3 of the slit length. On all the other targets, a 2$\times$3 step mapping observation was carried out: 
the three steps were chosen to be separated by three-quartes (for SL) or half (for LL) of the slit width in 
the dispersion direction and the two steps were separated by one-third of the slit in the spatial direction.
The exposure time at each nod position was 6\,second for all targets with the exception of V710~Tau which had a 
14\,second exposure time per nod.

We downloaded the low--resolution IRS data  that were processed with
the SSC pipeline S13.2.0. Our data reduction starts from the {\it droopres} intermediate data product 
and follows the steps outlined in detail in \citet{bouwman07}.
In brief,  we first subtracted the pairs of imaged spectra acquired along the spatial direction of the slit 
in order to correct for the background emission and stray--light. Then we replaced bad pixels
by interpolating over neighboring, good pixels. Although for our analysis we use only the SL data, we also extracted the LL part of the spectra for comparison with the MIPS 24\,\micron{} photometry where both datasets are available.  Spectra were extracted from the 
background--subtracted pixel--corrected images using a 6.0--pixel and 5.0--pixel fixed--width aperture in the
spatial direction for the SL (5.2--14\,\micron) and LL
(14---35\,\micron) modules, respectively. The low--level fringing at wavelengths $>$20\,\micron{} was removed using the irsfringe package \citep{lb03}.

Because peak--up images were not acquired for the majority of the sources, targets may not have been positioned
accurately at the center of the slit. We determined the actual position of each source during extraction by finding the peak emission of the wavelength--collapsed source profile. 
Once the spectra are extracted for each order, nod and cycle, we computed a mean spectrum for each order and assigned  as uncertainty at each wavelength the  1--sigma standard deviation of the distribution of the data points.

\begin{figure*}
\includegraphics[angle=90,scale=0.7]{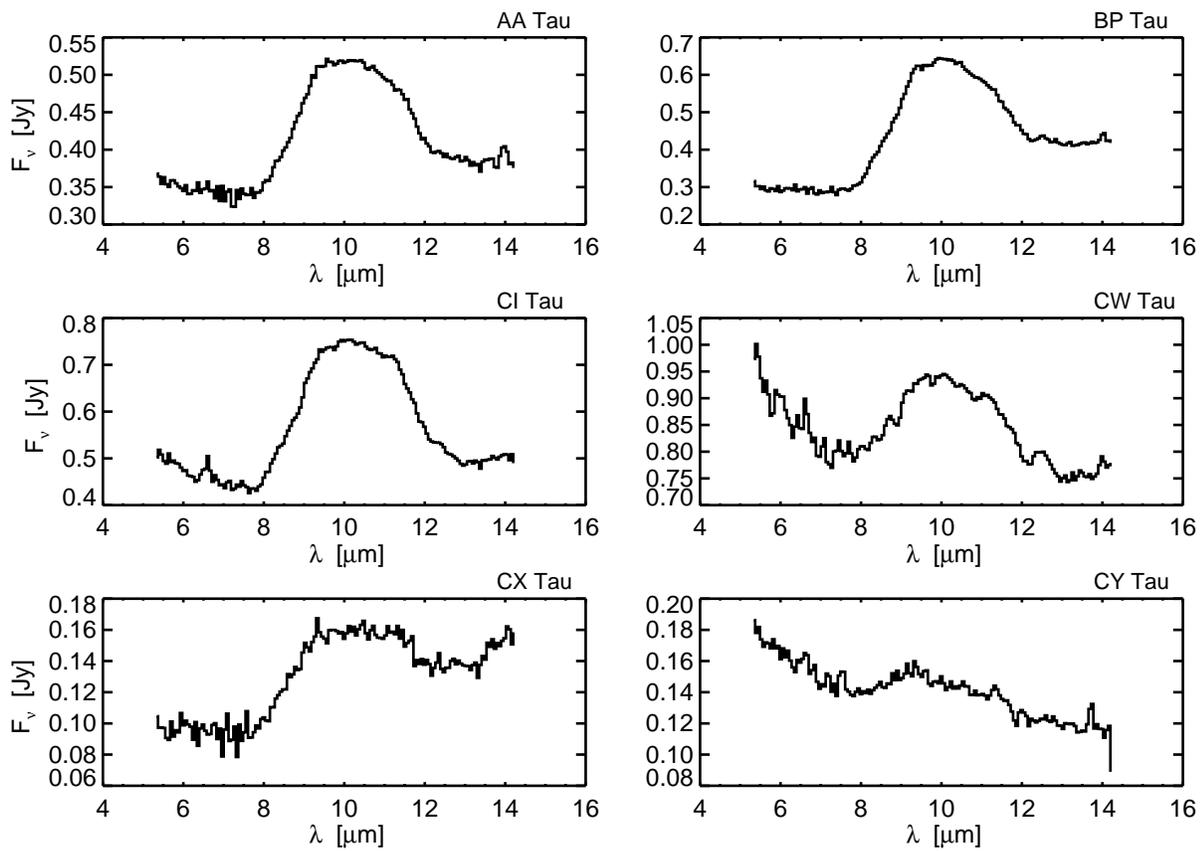}
\caption{Infrared spectra for the sample of single stars.\label{s1}}
\end{figure*}

\begin{figure*}
\includegraphics[angle=90,scale=0.7]{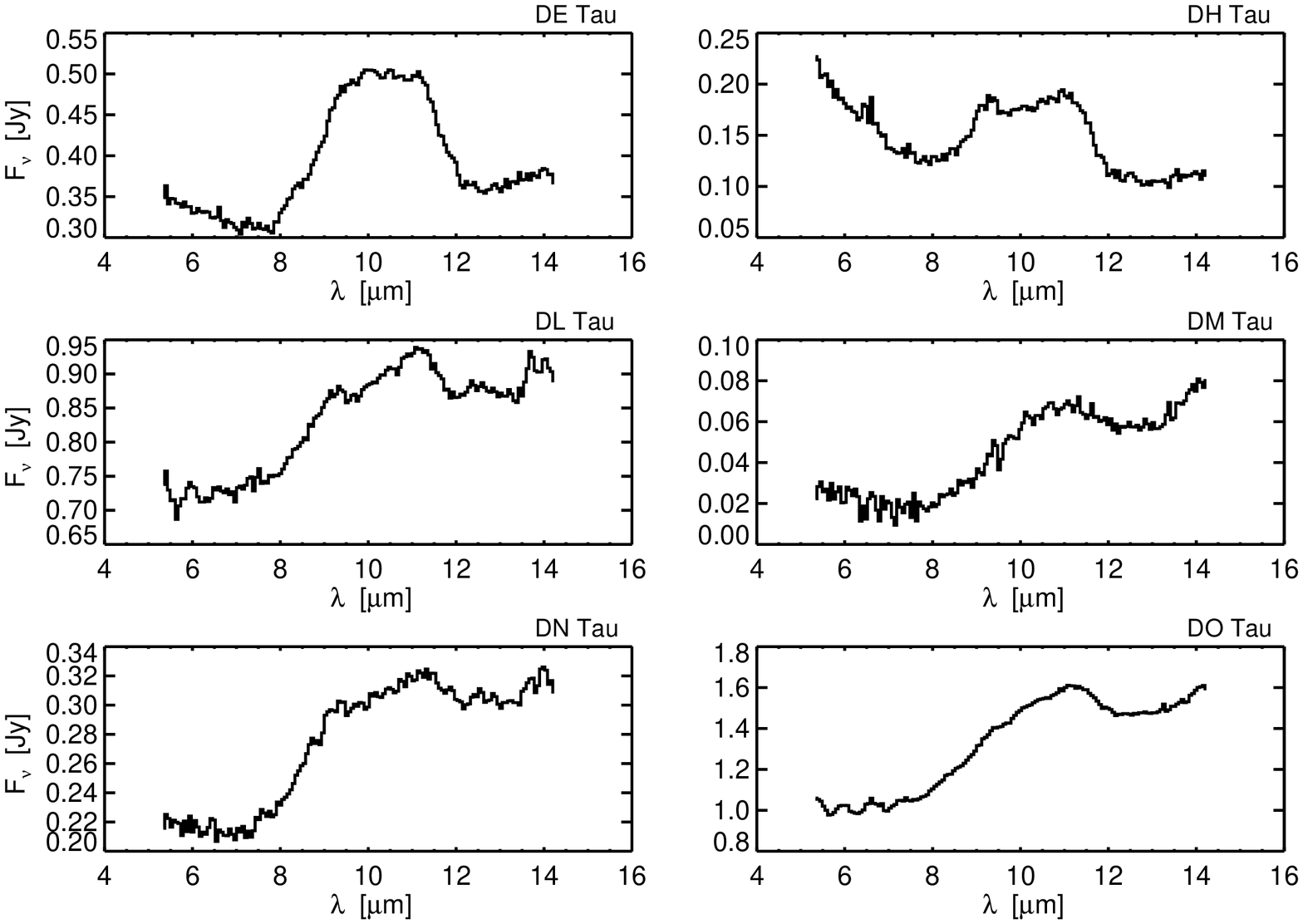}
\caption{Infrared spectra for the sample of single stars.\label{s2}}
\end{figure*}

\begin{figure*}
\includegraphics[angle=90,scale=0.7]{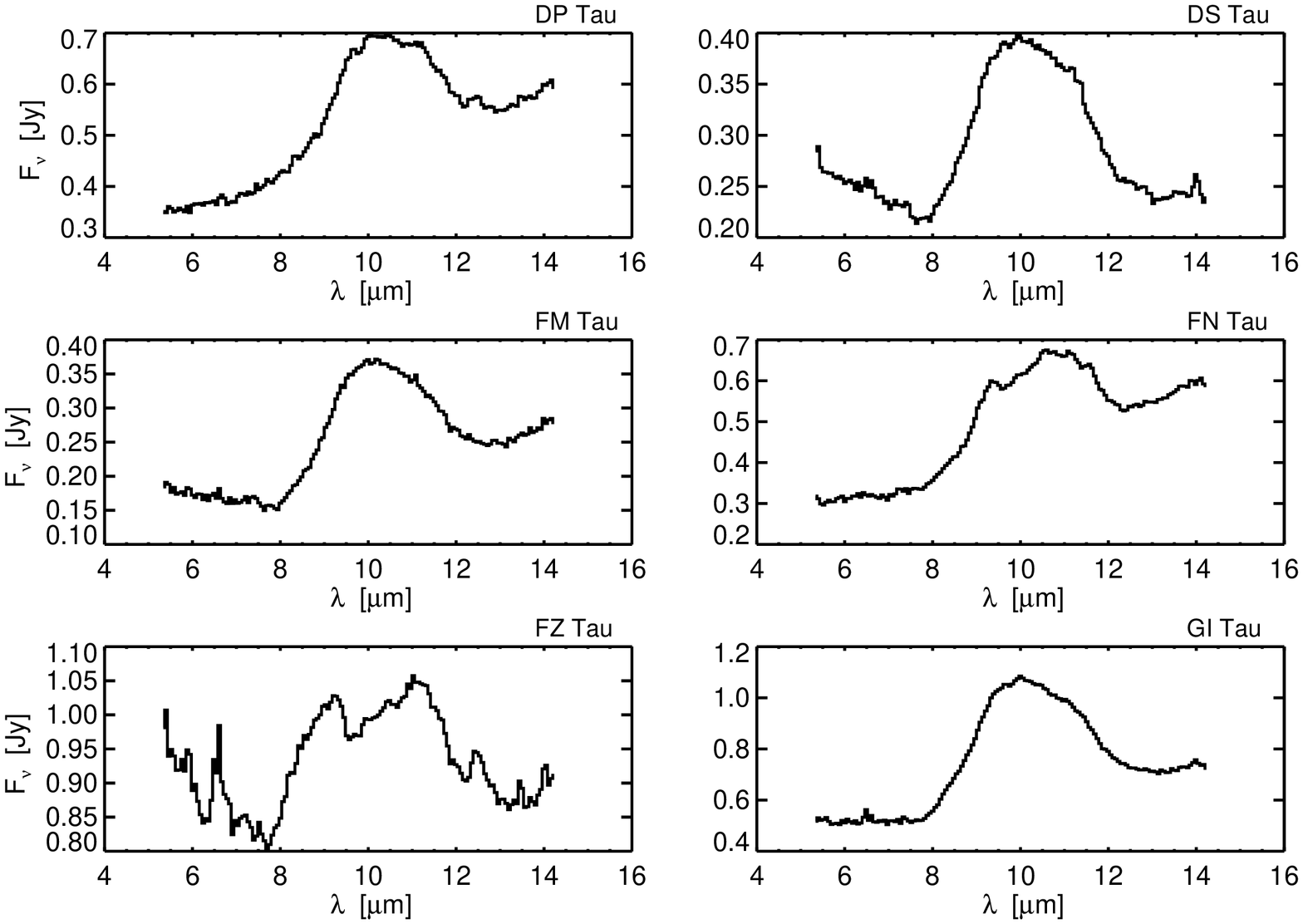}
\caption{Infrared spectra for the sample of single stars.\label{s3}}
\end{figure*}

\begin{figure*}
\includegraphics[angle=90,scale=0.7]{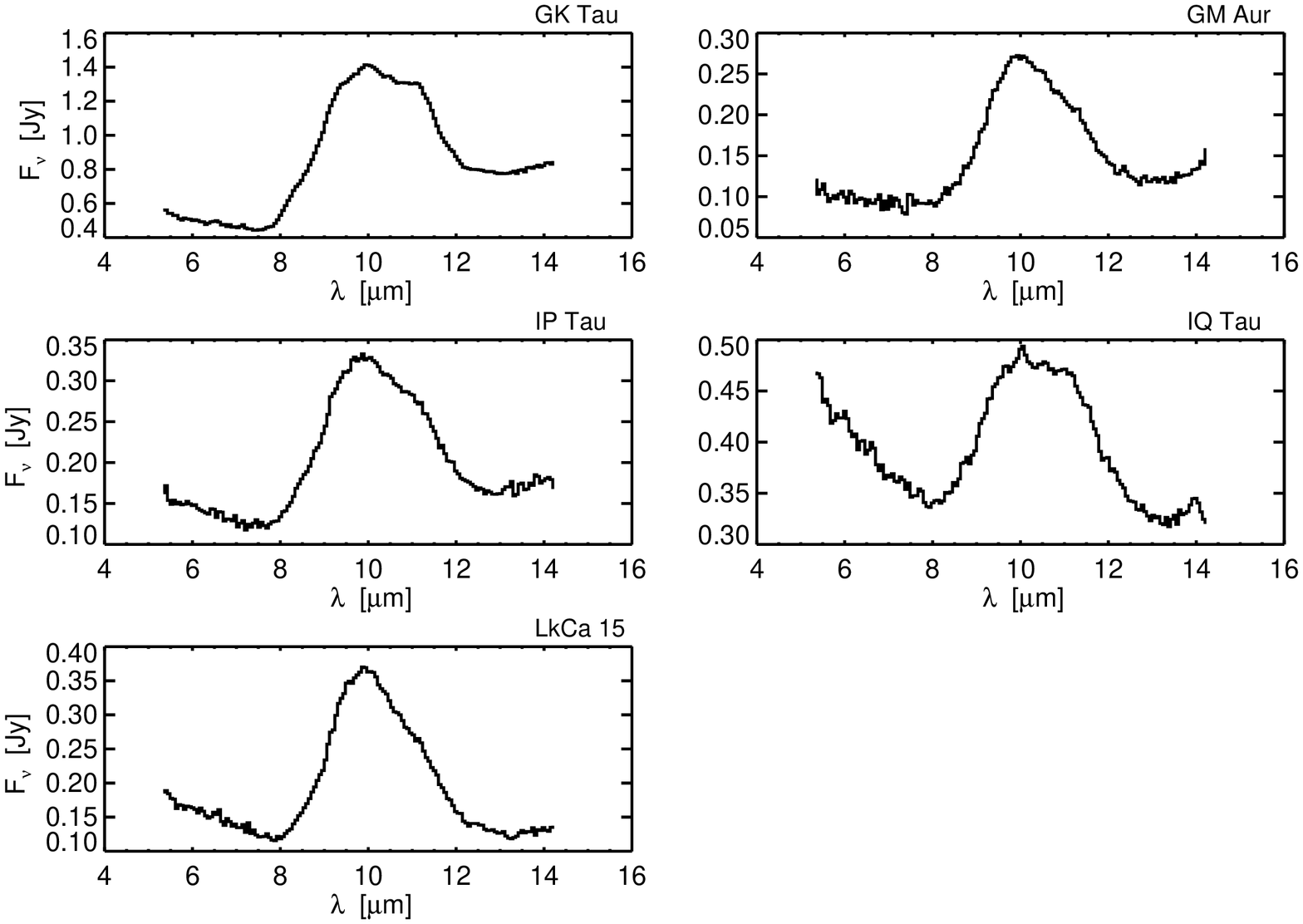}
\caption{Infrared spectra for the sample of single stars.\label{s4}}
\end{figure*}

\begin{figure*}
\includegraphics[angle=90,scale=0.7]{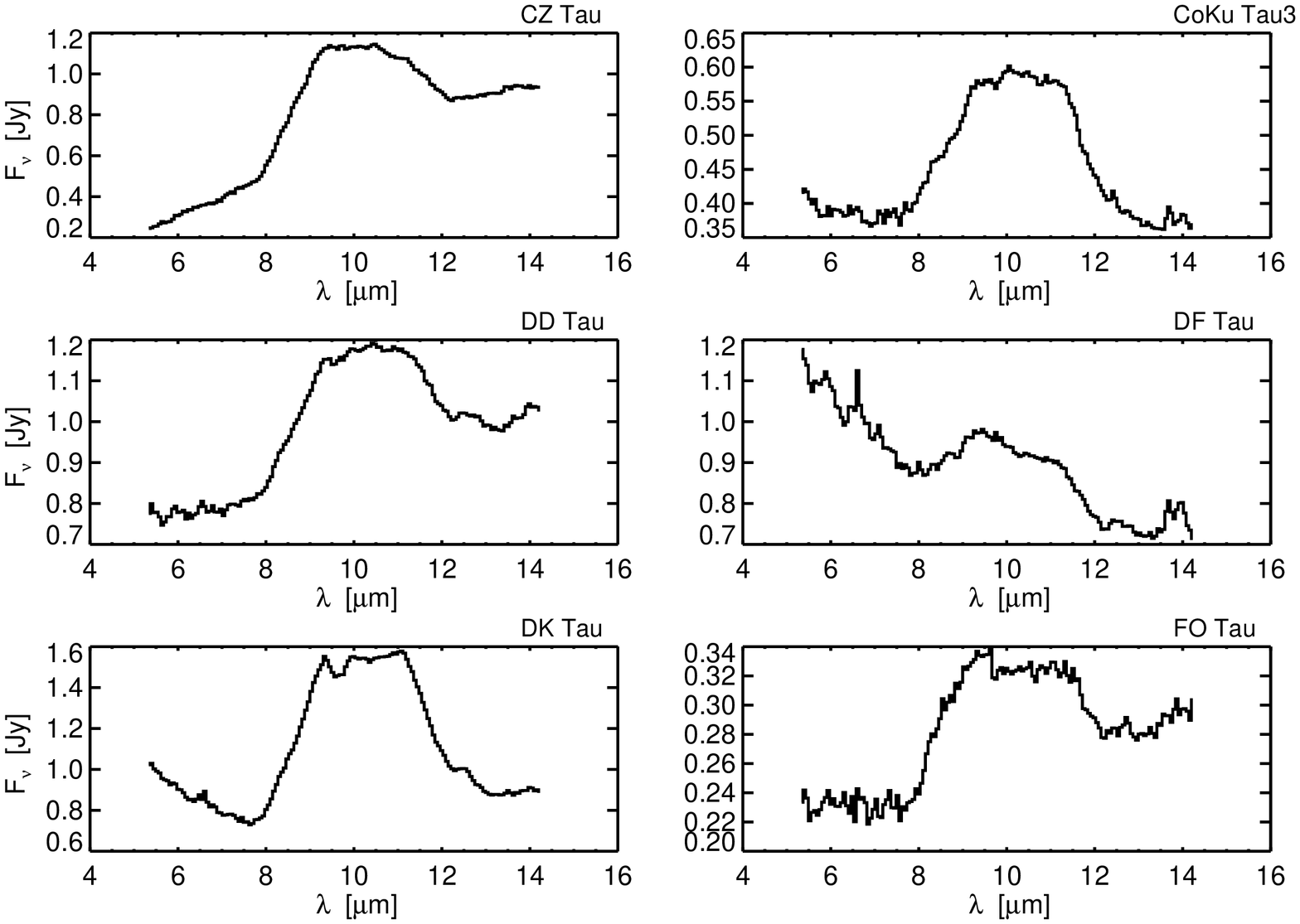}
\caption{Infrared spectra for the sample of binary stars.\label{b1}}
\end{figure*}

\begin{figure*}
\includegraphics[angle=90,scale=0.7]{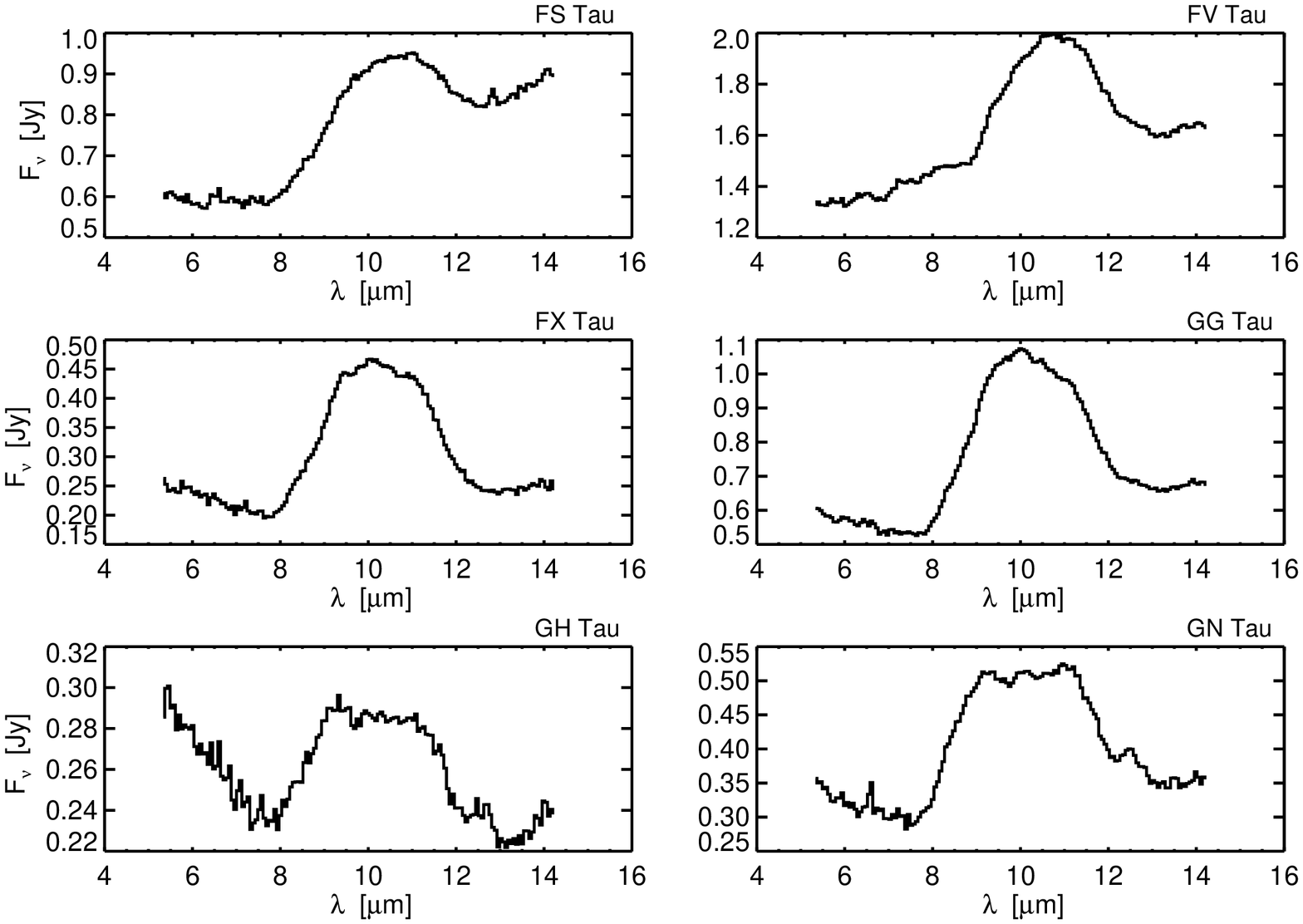}
\caption{Infrared spectra for the sample of binary stars.\label{b2}}
\end{figure*}

\begin{figure*}
\includegraphics[angle=90,scale=0.7]{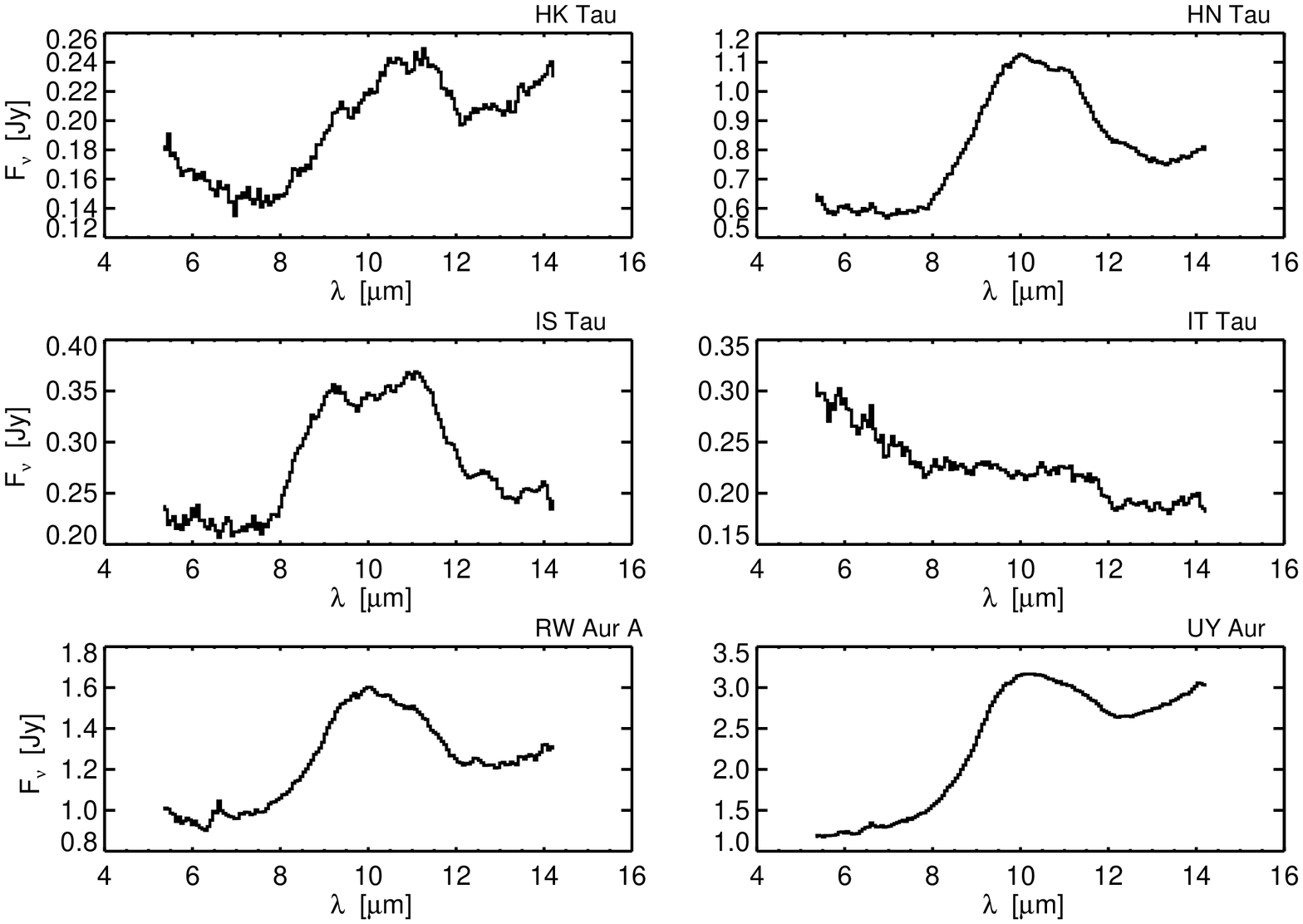}
\caption{Infrared spectra for the sample of binary stars.\label{b3}}
\end{figure*}

\begin{figure*}
\includegraphics[angle=90,scale=0.7]{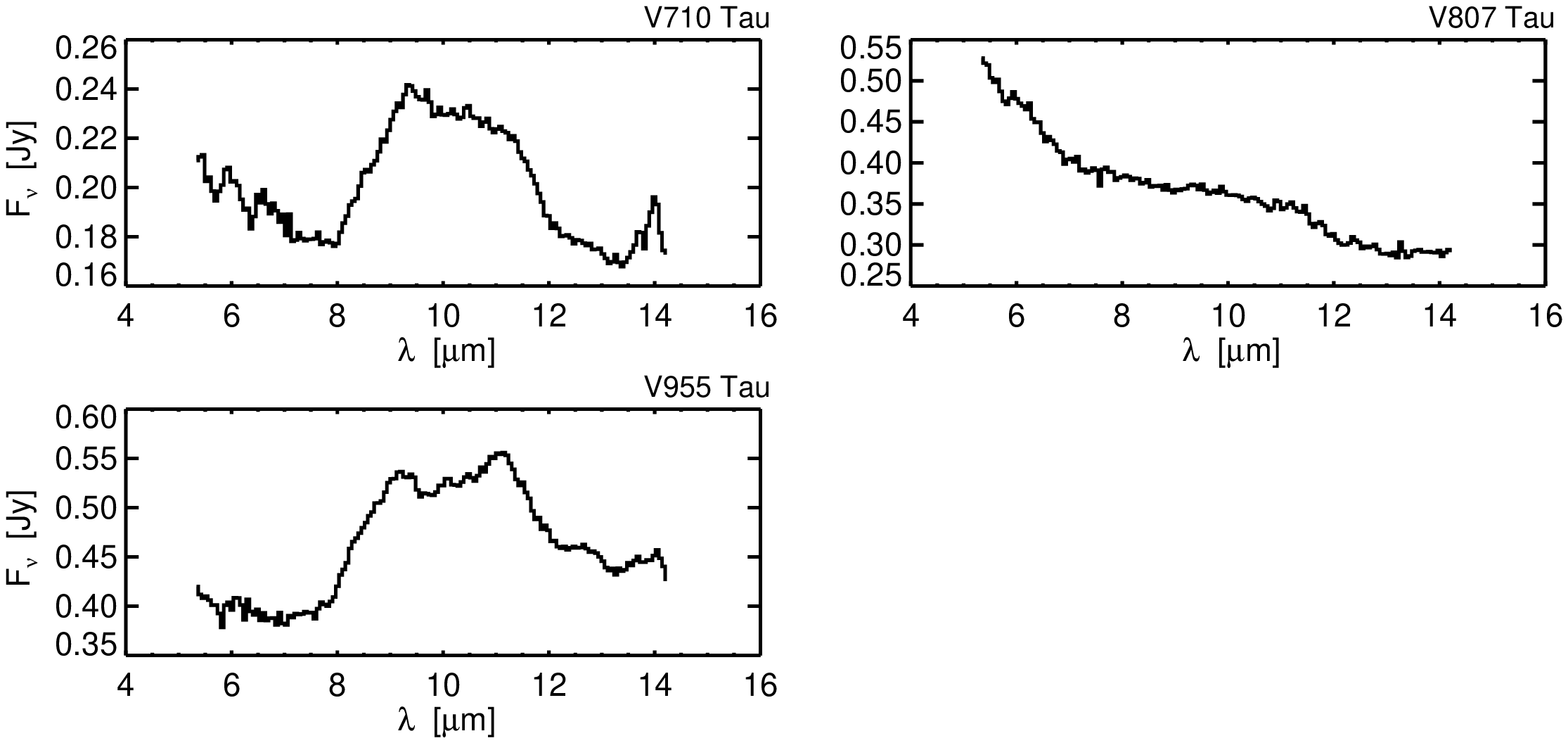}
\caption{Infrared spectra for the sample of binary stars.\label{b4}}
\end{figure*}

The absolute flux calibration was done using order--based spectral response functions created within the
{\it Formation and Evolution of Planetary Systems} (FEPS) Spitzer Legacy program \citep{hines05, meyer06,bouwman07}.  
The advantages of this spectral response function over the standard SSC bcd calibration are: 
a) the use of a larger number of calibrators from the FEPS 
program; b) the spectral response function is order and nod--position based; c) it is a 1D spectral response function 
allowing a better rejection of bad pixels. The estimated absolute flux calibration uncertainty is around 10\% \citep{bouwman07} and is propagated to the flux uncertainties assigned at each wavelength.
The resulting calibrated spectra are available in the electronic edition of the
Astrophysical Journal (Figs.~\ref{s1},~\ref{s2},~\ref{s3},~\ref{s4},~\ref{b1},~\ref{b2},~\ref{b3} and~\ref{b4}). 

We checked the absolute flux calibration of the IRS SL module using the published IRAC 8\,\micron{} photometry by \citet{hartmann05} and \citet{luhman06}, whose data have been acquired in two different campaigns, February--March 2004 and February 2005 respectively. Eleven sources have IRAC 8\,\micron{} magnitudes from both campaigns, while 5 more sources have only magnitudes from  February--March 2004 \citep{hartmann05}, and 14 more sources only  from February 2005 
\citep{luhman06}. The mean difference in the IRAC 8\,\micron{} fluxes for the 11 sources observed in both campaigns is 16\%, much larger than the IRAC absolute flux calibration accuracy of a few percent \citep{reach05}. The largest deviations are for FV~Tau, DL~Tau, and DO~Tau whose 8\,\micron{} IRAC fluxes decreased by factors of 33\%, 23\%, and 23\% respectively in a year baseline.
These differences are likely due to intrinsic stellar variability. The IRS 8\,\micron{} fluxes integrated over the IRAC 8\,\micron{} spectral response curve agree on average within $\sim$10\% of the published IRAC photometry, which is within the estimated
IRS flux calibration uncertainty. We also verified that our spectra very well agree with those published by \citet{furlan06} who adopted a different data reduction.

\begin{deluxetable}{cl c c c}
\tabletypesize{\tiny}
\tablecaption{Fluxes at 24\,\micron{} from MIPS images (third column, F24) or at 25\,\micron{} from IRS spectra (fourth column, F25).  The absolute photometric uncertainty is expected to be $\sim$10\%. This uncertainty includes  both the internal random and the absolute calibration uncertainty, see text for details.\label{T:phot24}}
\tablewidth{0pt} 					       
\tablehead{
\colhead{\#}&\colhead{Source}&\colhead{F24}&\colhead{F25}&\colhead{Comment}  \\
\colhead{} & \colhead{} & \colhead{[mJy]} & \colhead{[mJy]} &  \colhead{} }       
\startdata						       
       S1 & AA~Tau & 525    & ... &  \\		       
       S2 & BP~Tau & 666    & ... & \\  		       
       S3 & CI~Tau & 997     & ... &   \\  			       
       S4 & CW~Tau & 1445 & ... &   \\				       
       S5 & CX~Tau & 296	& ... &  \\  			       
       S6 & CY~Tau & 123	& ... &  \\
       S7 & DE~Tau & 677	& ... & \\				       
       S8 & DH~Tau & 316	& ... & \\				       
       S9 & DL~Tau & 965	& ... &    \\  			       
      S10 & DM~Tau & ... & 320 & no MIPS24 \\			       
      S11 & DN~Tau & 424 & ... &\\			       
      S12 & DO~Tau & ... & 2951 & MIPS24 saturated\\	       
      S13 & DP~Tau & 1281 & ... &\\	       
      S14 & DS~Tau & ... & 303 & no MIPS24 \\
      S15 & FM~Tau & 462 & ... & \\	      
      S16 & FN~Tau & 1047& ... & \\	      
       S17 & FZ~Tau & 1057& ... &  \\
       S18 & GI~Tau & 1006& ... & \\
       S19 & GK~Tau & ... & 1699  & MIPS24 saturated \\
       S20 & GM~Aur & 746 & ... &  \\	       
       S21 & IP~Tau & 277 & ... & \\
       S22 & IQ~Tau & 544& ... & \\
       S23 & LkCa~15& 398& ... &  \\	
	B1 & CoKu~Tau3 & 327 & ... &  \\
	B2 & CZ~Tau     & 1232 & ... &  \\
	B3 & DD~Tau 	& 1575 & ... & \\
	B4 & DF~Tau  	& 945 &   ... &\\
	B5 & DK~Tau  	& 1286 & ... & \\   
	B6 & FO~Tau  	& 529 &  ... & \\    
	B7 & FS~Tau  	& ... & 2112 & MIPS24 saturated\\
	B8 & FV~Tau  	& ... & 2344 & MIPS24 saturated\\
	B9 & FX~Tau  	& 420 &  ... & \\
        B10 & GG~Tau & ... & 1266 & no MIPS24\\
	B11 & GH~Tau 	& 379 & ... & \\
	B12 & GN~Tau 	& 536 & ... &   \\
	B13 & HK~Tau 	& 823 & ... &  \\
	B14 & HN~Tau  &  & 2988 & no MIPS24\\
	B15 & IS~Tau   & 228 & ... & \\
	B16 & IT~Tau   & 264 & ... &  \\
	B17 & RW~Aur & ...  & 2012 & no MIPS24\\
	B18 & UY~Aur  & ... & 5918 & no MIPS24\\
	B19 & V710~Tau  & 245 & ... &\\
	B20 & V807~Tau  & 476 & ... &\\
	B21 & V955~Tau  & 546 & ... & \\  
\enddata
\end{deluxetable}

\subsection{{\it Spitzer} 24\,\micron{} photometry}\label{S:mips24}
MIPS 24\,\micron{}  (hereafter MIPS24) data are available in the {\it Spitzer} archive for all but 6 of the targets we selected (specifically DM~Tau, DS~Tau, GG~Tau, HN~Tau, RW~Aur, and UY~Aur). The majority of the MIPS data have been acquired as part of the Taurus {\it Spitzer} Legacy Program (PI D. Padgett) in February--March 2005 (id=3584) and March 2007 (id=30816) using the MIPS scan map operational mode. V710~Tau and GM~Aur are not covered by these programs but have data from the c2d {\it Spitzer} Legacy Program (PI N. Evans, pid=173) and the GTO program respectively (PI G. Fazio, pid=37).
We downloaded the post--bcd MIPS 24\,\micron{} products processed through the SSC pipelines S14.4.0 or later. 
In the case of photometry mode (only for V710~Tau), the SSC product is an averaged and registered single image while in the case of scan maps (for all other sources), the product is a distortion--corrected mosaic image\footnote{http://ssc.spitzer.caltech.edu/mips/dh/mipsdatahandbook3.2.pdf}.
The 24\,\micron{} post--bcd images generated from pipeline versions later than S14 have improved flat field corrections and are suitable for photometry at an accuracy of 10\% (Sect. 9.1 of the MIPS Data Handbook v 3.2, link in footnote 3), sufficient for the purposes of this study. 

Aperture photometry was done using IDP3 \citep{idp3}. The centeroid
of the aperture was found by fitting a Gaussian to each source. 
We used an aperture radius of 6\farcs6 (2.7 pixels) and background annulus
from 20\arcsec{} to 32\arcsec. We opted for this intermediate aperture radius among those suggested by the Spitzer Science Center to include the emission from both binaries (as in the IRS spectra) and to exclude the emission from companions at $>$\,10\arcsec{} from some of the single stars (see, note to Table~\ref{T:single}).
To recover the total flux we then applied the proper aperture correction  of 1.648 \footnote{http://ssc.spitzer.caltech.edu/mips/apercorr/}.
DO~Tau, GK~Tau, FS~Tau, and FV~Tau have 24\,\micron{} fluxes $\ge$1.6\,Jy and are thus saturated in the 3\,sec exposures
acquired within the Taurus {\it Spitzer} Legacy Program. For these sources as well as for the 6 targets that lack MIPS24 data, we computed the 25\,\micron{} flux (integrated flux from 23.5 to 26.5\,\micron) from our SL 6 \,\micron{} flux and the n$_{6-25}$ spectral indices from Table~4 of \citet{furlan06}\footnote{most of these sources are so bright that were observed only with the IRS high--resolution module. Therefore we could not compute 24\,\micron{} fluxes from the LL data we reduced}. We will show in Sect.~\ref{S:flaring} that our results are not affected by the use of 25\,\micron{} fluxes instead of 24\,\micron{} fluxes for these 10 objects.
For the  sources that have both MIPS24 observations and LL data we find that our aperture photometry agrees within about 10\% with the IRS 24\,\micron{} flux integrated over the MIPS24 spectral response curve. Table~\ref{T:phot24} summarize the MIPS24 or the IRS25 fluxes for our sample of single and binary systems.

\section{Results}\label{S:results}

\begin{figure*}
\includegraphics[angle=90,scale=0.7]{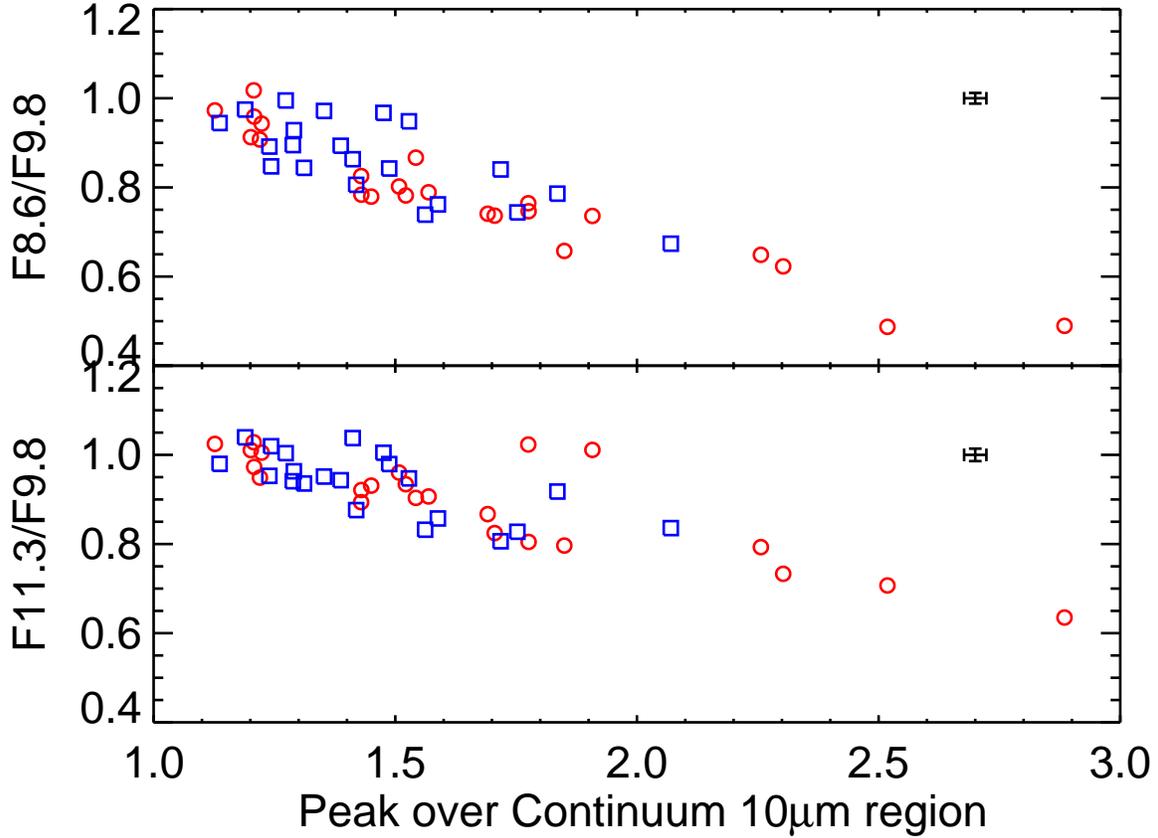}
\caption{Strengths and shapes of the 10\,\micron{} silicate emission features for the sample of single (red circles)
and binary (blue squares) systems. On the y--axis we plot the ratios of the normalized flux at 8.6 and 11.3\,\micron{} over 
9.8\,\micron , which are proxies for the degree of crystallinity. The x--axis gives the peak over continuum in the 10\,\micron{} region of normalized spectra, which indicates the amount of grain growth. Objects with processed dust are located on the top left of the diagram. Average error bars are shown on the right upper corner of each plot (see Sect.~\ref{Sect:ss} for details).\label{F:ss}}
\end{figure*}

\subsection{Strength  and Shape of the 10\,\micron{} Feature}\label{Sect:ss}
The 10\,\micron{} silicate emission feature traces silicate dust in the optically thin layer of circumstellar disks out to about 1\,AU from a star with solar luminosity (e.g. \citealt{kessler07}). Because the smaller the grain the longer its timescale to settle to the disk midplane, the 10\,\micron{} feature probes the population of the smallest grains in a disk, up to a few micron in size.
The strength and shape of the feature bear information on the amount of processing that dust grains have undergone in the disk. In particular, the flux ratio of normalized spectra at 11.3 (or 8.6) over 9.8\,\micron{} can be used as a proxy for the degree of crystallization while the peak--to--continuum ratio can be used as a proxy for grain growth (e.g. \citealt{bouwman01,van05,apai05,bouwman07}). We use these band strengths to search for differences in the dust processing of single and binary systems.

We first fit a third--order  polynomial\footnote{A third order polynomial gives smaller residuals in the spectrum minus fitted continuum than a first, a second, or a fourth order polynomial. We also verified that the results discussed in this section do not change when using lower or higher order polynomials to fit the continuum.} to the spectral data outside the 10\,\micron{} silicate emission feature (between 6 and 8\,\micron{} and between 12 and 14\,\micron) and then
normalize our spectra to the fitted continuum. This normalization ensures that the shape of the spectral features 
remains identical to the original one (e.g. \citealt{van05}). Fig.~\ref{F:ss} shows the band strengths for our sample of single (red circles) and multiple (blue squares) systems. The flux densities at 8.6, 9.8, and 11.3\,\micron{} are the mean values of flux densities within $\pm$0.1\,\micron{}  of these wavelengths. To analyze the uncertainties we followed a Monte Carlo approach. For each disk we added a normally distributed noise to the spectrum and computed the peak--over--continuum and flux ratios 500 times. Their standard deviation gives the uncertainty on the  peak--over--continuum and flux ratios for each target.  The error bars in Fig.~\ref{F:ss} are the average uncertainties. The simulated noise had two components: i) a normally distributed noise at each wavelength with an amplitude equal to the flux uncertainty at that wavelength; ii) a random calibration uncertainty (equal at all wavelengths) with an amplitude of 10\%.

\begin{figure*}
\includegraphics[angle=90,scale=0.7]{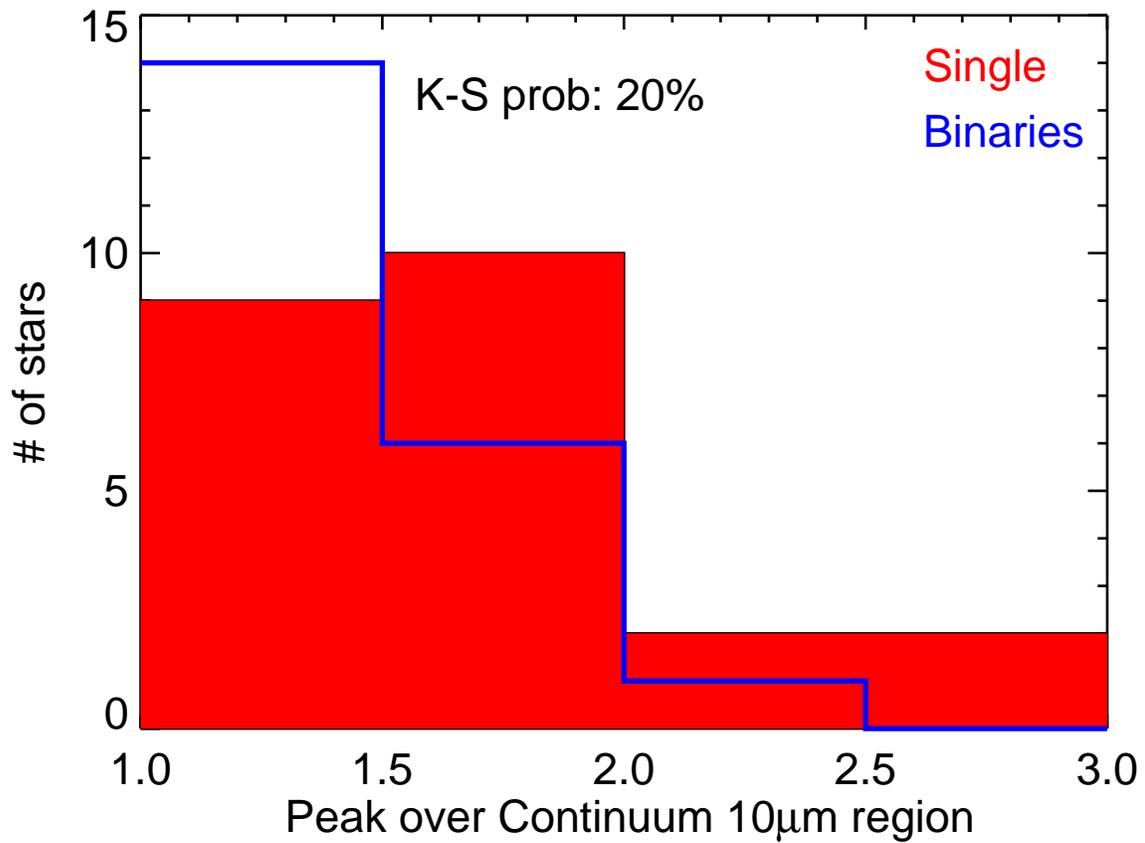}
\caption{Histogram of the 10\,\micron{} feature strengths  for the single (red) and binary (blue) systems. 
The K--S test indicates that the two distributions are consistent with having been drawn from the same parent population.  This suggests that the growth of grains up to a few micron in size is not affected by
the presence of a medium--separation stellar companion.  \label{F:ratio}}
\end{figure*}

\begin{figure*}
\includegraphics[angle=90,scale=0.7]{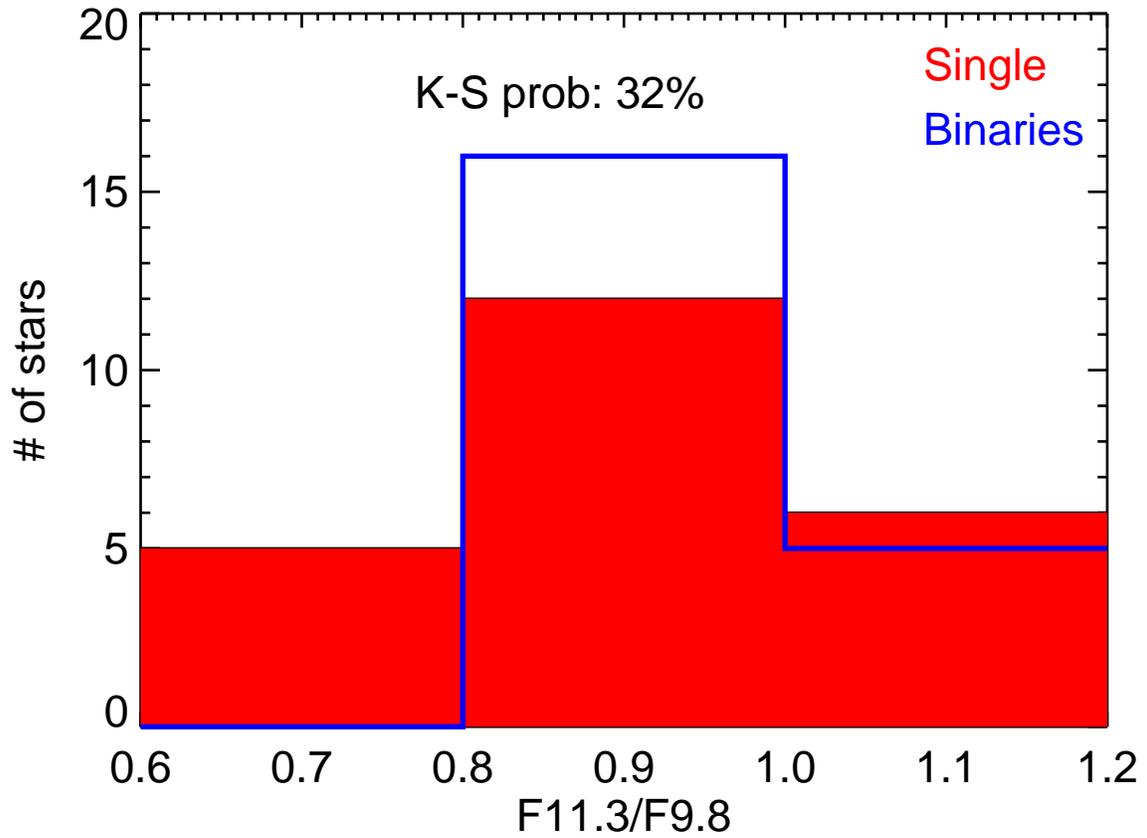}
\caption{Histogram of the flux ratios at 11.3 over 9.8\,\micron{}  for the single (red) and binary (blue) systems.  
The K--S test does not indicate that the two distributions differ statistically. This suggests that crystalline processing is not influenced by the presence of stellar companions at tens of AU. \label{F:peak}}
\end{figure*}

\begin{figure*}
\includegraphics[angle=90,scale=0.7]{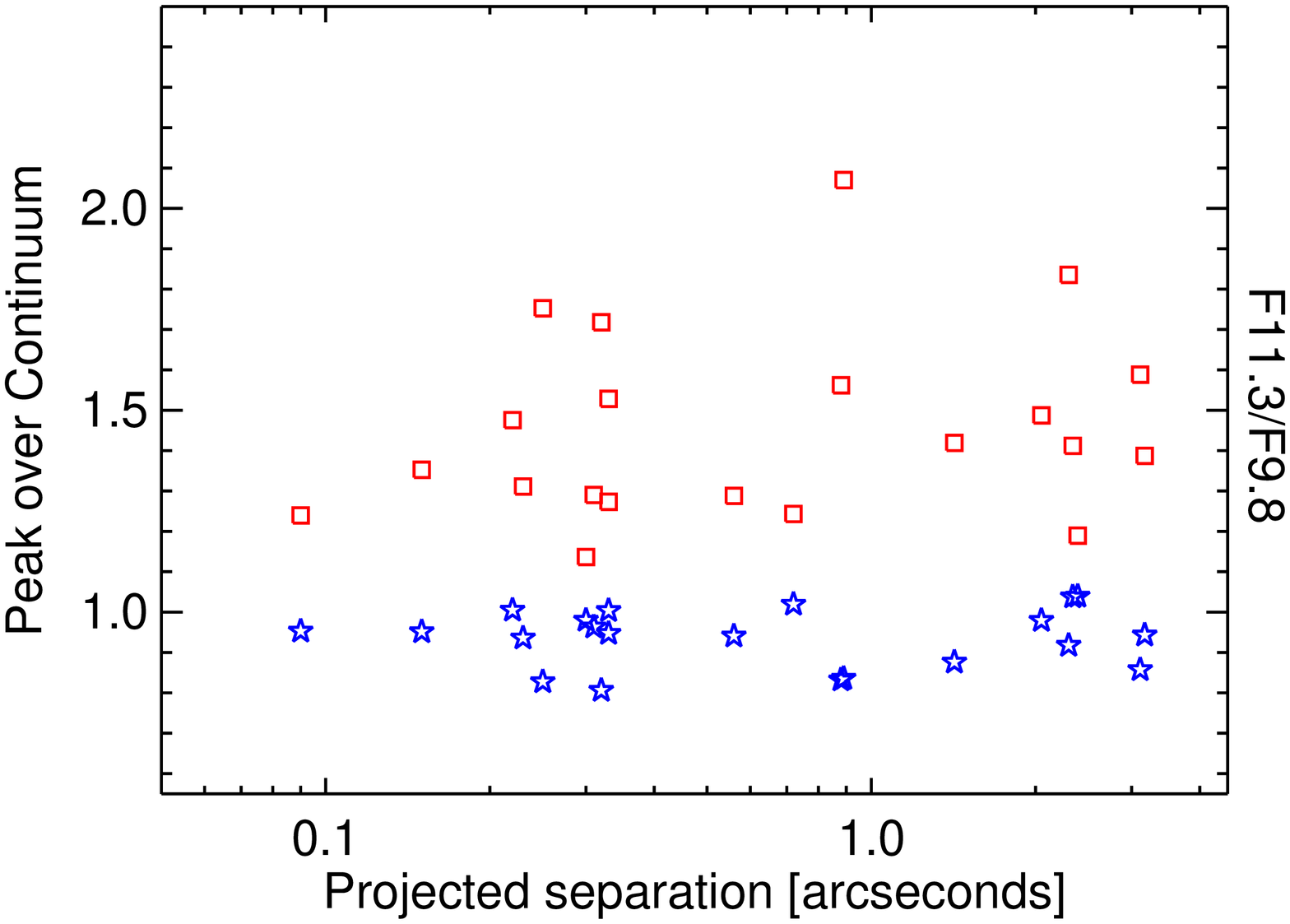}
\caption{Strength (squares) and shape (stars) of the 10\,\micron{} silicate emission feature versus projected binary separation. The peak--over--continuum compared to the projected separation (squares) gives
a Kendall's $\tau$=0.1 and $P$=0.4 suggesting that these variables are not correlated. Similarly, the F11.3/F9.8 flux ratio is uncorrelated with the binary projected separation (stars) as indicated by a Kendall's $\tau$=0.02 and 
$P$=0.9.  \label{F:separ}}
\end{figure*}

Objects on the bottom right side of Fig.~\ref{F:ss}  have 10\,\micron{} emission features similar to the ISM absorption feature, thus  their dust grains have experienced little processing. Conversely, objects on the top left side of 
the plot have disks with larger grains (micron in size) and crystals. 
There is an obvious anticorrelation between  the strength of the 10\,\micron{} silicate emission feature and the F11.3/F9.8 (or F8.6/F9.8) flux ratios, which is apparent not only in Fig.~\ref{F:ss} but also in other similar plots in the literature (e.g. \citealt{van05,apai05}). We use the Kendall's $\tau$ test to measure the degree of anticorrelation. 
This nonparametric test uses the relative rank ordering of pairs of data to compute two values (see e.g. \citealt{press03}): the rank correlation coefficient $\tau$  which runs between -1 (complete rank reversal) and 1 (complete rank agreement), and the two--sided probability $P$ that the variables are uncorrelated ($\tau=0$).
For the single stars in Fig.~\ref{F:ss} (circles), Kendall's $\tau=-0.8$ and $P=6\times10^{-8}$ for the F8.6/F9.8
flux ratio versus the peak--over--continuum while $\tau=-0.6$ and $P=2\times10^{-5}$ for the F11.3/F9.8 flux ratio versus the peak--over--continuum. For the binary stars in Fig.~\ref{F:ss} (squares),  Kendall's $\tau=-0.5$ and 
$P<0.001$ for the F8.6/F9.8 and the F11.3/F9.8 flux ratios versus the peak--over--continuum.  We interpret these
trends as confirmation that there is a significant anticorrelation between the strength of the silicate emission features and the flux ratios presented in Fig.~\ref{F:ss}.  
Therefore, both the peak--over--continuum and the F11.3/F9.8 (or F8.6/F9.8) flux ratios provide an analogue measurement  of dust processing in circumstellar disks.


The main question we want to answer in this paper is whether the population of disks around single stars statistically differ in terms of dust processing from the population of disks around binaries. In other words, do either of the  two populations have more processed dust in their disks than the other?
To answer this question we apply the Kolmogorov--Smirnov (hereafter K--S) test (see, e.g. \citealt{press03}) and the Mann--Whitney U (hereafter MWU) test (see, e.g. \citealt{ronald}) on the peak--over--continuum and the flux ratios (F11.3/F9.8 and F8.6/F9.8) presented in Fig.~\ref{F:ss}.
The K--S and the MWU tests are both non--parametric tests with the null hypothesis that two samples are drawn from a single population and work on unbinned data.
Because the K--S test gives large probabilities of 0.2 for the peak--over--continuum of single and binaries and 0.3 for their 
F11.3/F9.8 (and F8.6/F9.8) flux ratios, we conclude that both samples could have been drawn from the same distribution. Similarly the MWU test does not reveal any statistically significant difference in the degree of grain growth and crystallinity of the single and binary samples: the probability that both distributions were drawn from the same parent population for the peak--over continuum is 0.08 while for the flux ratios F11.3/F9.8 (and F8.6/F9.8) is 0.1.
Figs.~ \ref{F:ratio} and \ref{F:peak} provide a visualization of the peak--over continuum and F11.3/F9.8 flux ratio distributions for the samples of disks around single and binary stars.
Further supporting our conclusion, Fig.~\ref{F:separ} shows no clear trend of the strength (or shape) of the 10\,\micron{} feature with the projected separation of the companion. Kendall's $\tau$ are close to 0 and $P$ are large for the rank ordering of peak--over-continuum (or F11.3/F9.8) compared to the  projected separation (see caption of Fig.~\ref{F:separ}) suggesting that the variables are uncorrelated.

\begin{figure*}
\includegraphics[angle=90,scale=0.7]{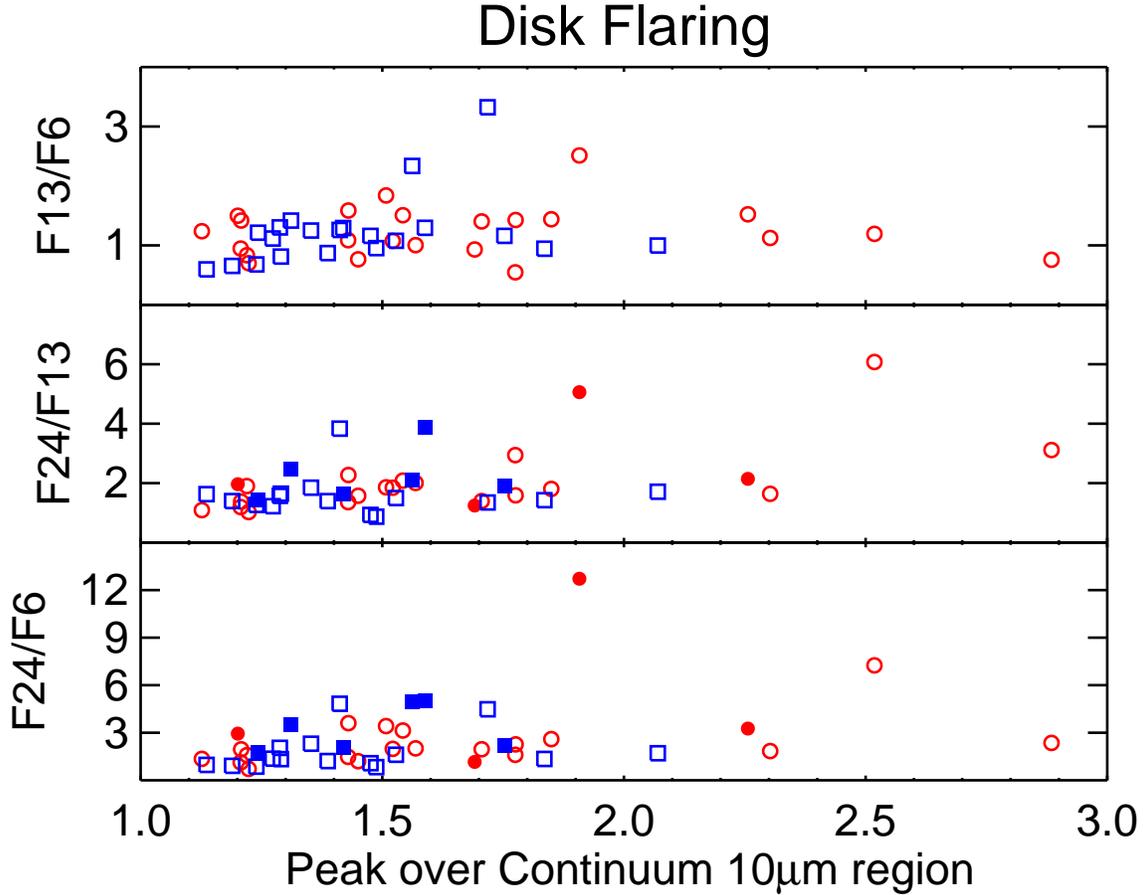}
\caption{Flux ratios at 13.25 over 5.7\,\micron{} (top), at 24.0  over 13.25\,\micron{} (middle), and at  24.0 over 5.7\,\micron{} (bottom) versus the peak--over--continuum in the 10\,\micron{} feature. These ratios are a proxy for the disk flaring; more flaring is indicated by larger ratios in the figure.  Squares represent binaries, while circles indicate single stars. Filled symbols are for those sources that do not have MIPS24  photometry or are saturated in the MIPS exposures. For these sources we plot the IRS25 flux computed as described in Sect.~\ref{S:mips24}. Both single and binary systems have a large variety of disk structures.\label{F:flaring1}}
\end{figure*}

\begin{figure*}
\includegraphics[angle=90,scale=0.7]{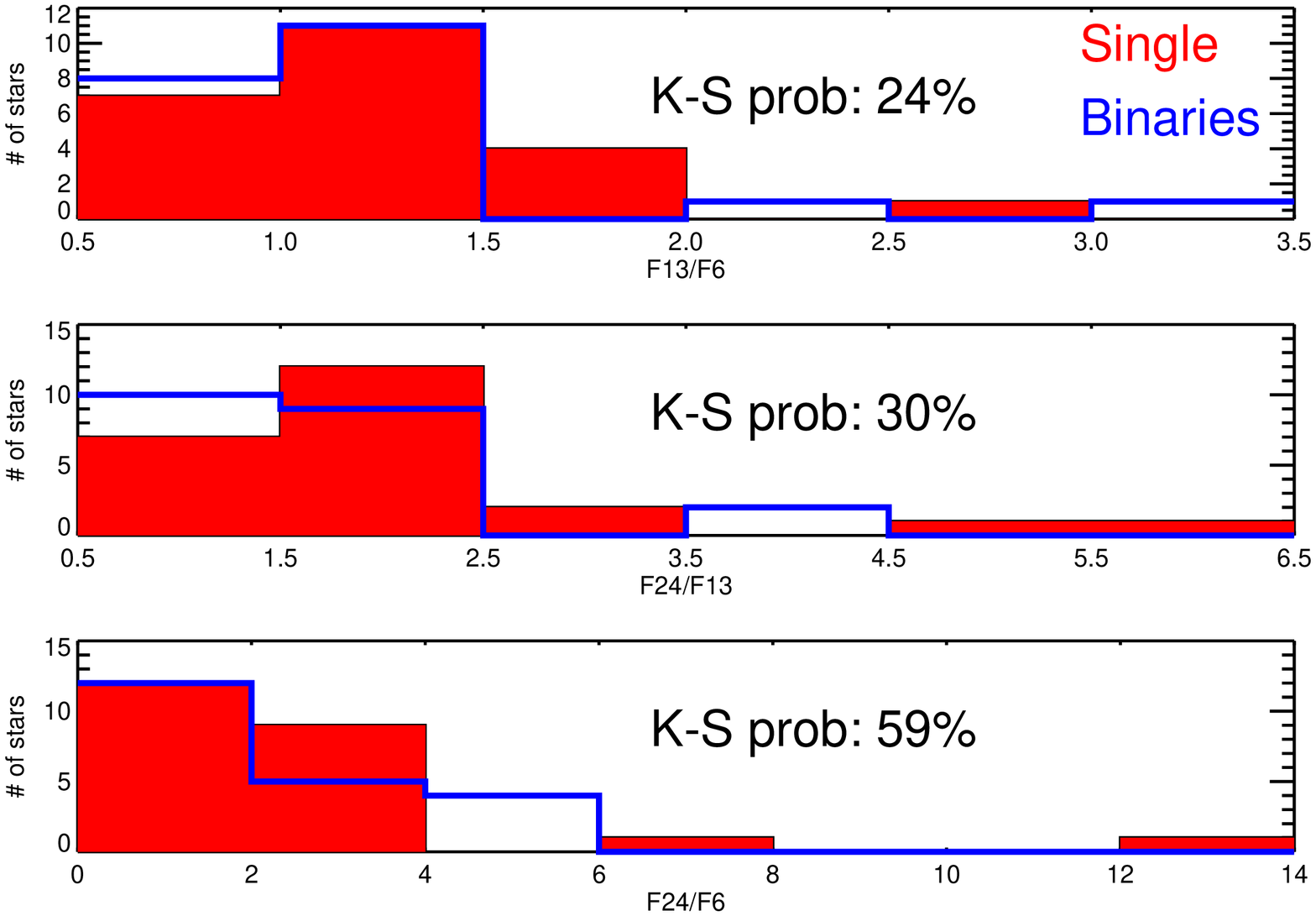}
\caption{Histograms for the flux ratios at 13.25 over 5.7\,\micron{} (top), at 24.0  over 13.25\,\micron{} (middle),
and at  24.0 over 5.7\,\micron{} (bottom) for the single (red) and binary (blue) systems. The distributions of flux
ratios (disk flaring) do not differ statistically according to the K--S and the MWU tests. MWU probabilities are
0.22,  0.13, and 0.18 for the F13/F6, F24/F13, and F24/F6 respectively. \label{F:flaring2}}
\end{figure*}
\subsection{Dust settling}\label{S:flaring}
Coagulation models show that if the dust grains in the upper layers of a flared disk 
become sufficiently large, they will gravitationally settle towards the mid--plane of the disk, 
resulting in a flattened disk geometry (e.g. \citealt{rainer04,nomura06}). Thus a way to search for grain growth is to evaluate the flaring of circumstellar disks. While the shape of the 10\,\micron{} feature is sensitive to the presence of grains of a few microns in size in the disk upper layer the disk flaring should probe the overall grain population of larger grains (e.g. \citealt{dd04}).
To evaluate the disk flaring we use the ratio of fluxes at two different mid--infrared wavelengths.
This procedure is justified by two facts: a) the continuum flux emitted from the surface layer of the disk becomes 
proportional to the disk flaring at radial distances from solar--type stars of 0.4\,AU or larger \citep{cg97}  and b) these radial distances are probed by mid--infrared observations. 

We integrate the flux of our SL infrared spectra in two wavelength bands one shortward and one longward of the 10\,\micron{} silicate emission feature: 5.4--6.0\,\micron{} (central wavelength 5.7\,\micron) and 12.5--14.0\,\micron{} (central wavelength 13.25\,\micron). In addition, we use the MIPS24 photometry (or the IRS25 flux when MIPS is not available, see Table~\ref{T:phot24}) to trace the disk flaring out to a few AU from the central star. The calculated flux ratios are shown in Fig.~\ref{F:flaring1} and histograms are presented in Fig.~\ref{F:flaring2}. Larger flaring is indicated by higher ratios of long--wavelength continuum to short--wavelength continuum flux from the dust disk. These plots show that both single and binary systems have a large variety of disk structures with no preference for any structure in the two samples. The K--S and the MWU probabilities that the
the flux ratio distributions for single and binary stars come from the same parent population are higher than 0.1 confirming that 
the distributions of disk flaring for the two samples are not statistically different (the same result holds when excluding the sources with IRS25 fluxes).

\section{Discussion}\label{S:discussion} 
Our study shows a large diversity in the 10\,\micron{} silicate emission features and SED slopes of T~Tauri disks. We found that neither the dust processing nor the disk flaring correlates with the multiplicity of the sources. 
These results are particularly interesting for two aspects that will be discussed in the following. 

\subsection{Medium--separation binaries and planet formation}
A stellar companion induces tidal forces in a disk that become particularly strong at resonance points. Resonant interactions result in the excitation of density waves that can truncate a disk and act to modify the binary eccentricity (see e.g. \citealt{lubow00} for a review). Theoretical calculations of binary--disk interactions predict that circumstellar disks will be truncated at 0.2--0.5 times the binary semimajor axis $a$, with the exact values depending on eccentricity, mass ratio, and disk viscosity \citep{art94}. These theoretical expectations are supported by millimeter observations of  binaries tracing the optically thin dust emission and thus the total disk mass in the system. There is evidence for a diminished millimeter flux (hence disk mass) among the 1--100\,AU binaries in comparison to wider binaries or single stars (e.g. \citealt{mathieu00} for a review). This result is qualitatively consistent with the circumstellar disks of medium--separation binaries being tidally truncated at 0.2--0.5$a$. Two--thirds of our sources have stellar companions between 0.1\arcsec--1\arcsec{}, with a mean projected separation of 0.4\arcsec{} or 56\,AU at the distance of Taurus. Therefore, the typical truncation radius for disks in our sample is  $>$11--28\,AU, well outside the location of Jupiter and Saturn in our Solar System. Even in other systems these outer radii are found to be devoid of giant planets  (e.g. \citealt{kasper07}). This fact suggests that the formation of terrestrial and giant planets may proceed undisturbed in disks around medium--separation binaries even if these disks are constrained in size.

Early investigations of young TTSs found no significant difference in the frequency of near-- and mid--infrared excess emission between single and binary star systems (e.g. \citealt{simonprato95,jensen96}). With the 60\micron{} IRAS flux probing dust $\la$10\,AU from the central star, these  measurements demonstrate that binary systems as often have disks as single stars do. 
Recently \citet{monin07} analyzed the separation distributions of binaries with and without disks and found no  statistical difference. Since most of their binaries have projected separations $>$20\,AU, their result shows that medium--and wide-- separation binaries do not have a significant effect on the circumstellar disk lifetime.
Our work indicates that these disks also evolve in a similar way. The extent of dust processing in the disk surface layer and the degree of dust settling in binary disk systems do not statistically differ from those in disks around single stars. This suggests that the first few Myr of disk evolution in the terrestrial (and maybe out to the giant) planet--forming region are not affected by medium--separation stellar companions. Whether the disk evolution proceeds undisturbed for tens of millions of years until planets are fully formed cannot yet be assessed observationally. \citet{bouwman06} estimate a mean disk dispersion timescale of $\sim$5\,Myr for close ($\le$4AU) binaries in contrast to a timescale of $\approx$9\,Myr for single star systems. They argue that the time available to form planets in close binary systems is considerably shorter than that in disks around single stars, which may inhibit planet formation.  The only two medium--separation binaries in their sample hint for a disk dispersal timescale comparable to that of single stars suggesting a similar disk evolution for single and medium--separation binary systems over the first $\sim$\,10\,Myr.

Exoplanet surveys offer us a glimpse into the frequency and properties of giant planets in multiple star systems. Recently \citet{eu07} reported 42 planets orbiting binary  and multiple stars (see, their Table 1). \citet{bondes07} analyze a subsample of radial velocity planet host stars with uniform planet detectability and  demonstrate that the overall frequency of giant planets in binaries is not statistically different from that of planets in single stars. However, they find indications for a lower frequency of radial velocity planets in the subgroup of close-- and medium--separation binaries ($<\,50-100$\,AU). In a complementary study, \citet{desbar07} find that the mass distribution of planets in binaries  with separations $<\,300-500$\,AU is statistically different from that around wider binaries and single stars: Massive planets in short--period orbits are found predominantly around close-- and medium--separation binaries.  
Taken together, the results from the frequency and properties of exoplanets suggest that 
a stellar companion with separation less than a few hundred AU affects giant planet formation and/or the subsequent  migration. Numerical simulations seem to support this notion. \citet{kley00} shows that a fairly eccentric ($e_{\rm bin}=0.5$) stellar companion at 50--100\,AU enhances the growth rate of a Jupiter mass planet embedded in a circumstellar disk and makes its inward migration more rapid. Recently, \citet{kn07} confirm these trends by following the evolution of a 30\,M$_\earth$ protoplanet in a disk truncated by a stellar companion at 18.5\,AU and $e_{\rm bin}=0.36$, like the $\gamma$~Cep binary system.
Our study shows that the early evolution of protoplanetary disks surrounding binary stars is similar to that in single stars indicating that that the differences in the exoplanet properties arise in the later stages of their formation and/or migration.  

Whether terrestrial planet formation is also affected by medium--separation binaries cannot be yet addressed observationally. Our study shows that the initial dust processing is not impacted by the presence of a stellar companion. Based on the fact that the build--up of planetesimals as large as the $\sim$500--km Vesta has occurred in the first 3.8$\pm$1.3~Myr of the Solar nebula \citep{kleine02}, it is reasonable to speculate on the basis of our study that the formation of planetesimals in binary and single systems proceed along, if not on identical avenues.  Another indication supporting this suggestion comes from the finding of a similar incidence of debris disks in Gyr--old single and binary stars \citep{trilling06}. If the debris dust is produced by colliding asteroids, then the similar  rate of debris dust in binaries implies that planetesimal formation is not inhibited by the presence of stellar companions. Recent simulations of the later stages of terrestrial planet formation show that rocky planets can form in a wide variety of binary systems \citep{quinta07}. The binary periastron is the most important parameter in limiting the number of forming planets and their range of orbits. \citet{quinta07} show that binaries with periastron $\ga$10\,AU, comprising most of the medium--separation binaries investigated in this paper, can form terrestrial planets over the entire range of orbits allowed for single stars. As a result more than 50\% of the binary systems in the Milky Way \citep{dm91} are wide enough to allow the formation of  Earth--like planets.

\subsection{The diversity in silicate features and SEDs}
Although  small sample statistics suggested a correlation between stellar multiplicity and initial dust processing \citep{meeus03,sterzik04,sic07}, our study demonstrates that  medium--separation stellar companions do not appreciably affect the growth and crystallization of dust in circumstellar disks.
Given the criteria applied to select our samples, we can also exclude that age, spectral type, and stellar environment can account for the large variety of observed silicate emission features and SED slopes in our study. 
There may be several other factors contributing to this diversity that will be fully explored in an upcoming contribution. In the following we briefly mention two of them:

Turbulence in circumstellar disks not only drives the accretion of gas onto the central star but also replenishes the disk atmosphere with more grains that can be larger in size. If the grains inferred from the 10\,\micron{} silicate emission feature reflect the level of disk turbulence, the strength of the features should depend on the stellar accretion rates.
\citet{sic07} note that stars with strong features tend to have large accretion rates in their sample of
several Myr old intermediate-- and low--mass stars. This trend may be the result of turbulence determining the grain population in the disk atmosphere.  Alternatively, the trend could be due to the more massive stars (that have typically larger accretion rates) in their sample heating larger disk area and thus producing stronger silicate emission features (see, e.g.  \citealt{kessler07}).  The tentative correlation seen in the sample of \citet{sic07}  needs to be confirmed using  a larger and more homogeneous sample of stars with well--determined accretion rates. 


Different initial conditions for the collapsing cores  may also leave their imprints on the formation and evolution of circumstellar disks. This possibility has been explored by \citet{dullemond06} to explain crystallization of dust grains in the early stages of disk evolution. In their model the level of  crystallinity depends crucially on the rotation rate of the collapsing cloud core because this determines the radius at which the infalling matter reaches the disk: rapidly rotating clouds would evolve into disks with low crystallinity, while slowly rotating clouds into disks with high crystallinity. 

\section{Summary}\label{S:summary} 
In this paper we explored the effect of a stellar companion on the initial growth and settling of dust grains in circumstellar disks. We constructed two large samples of disks around single and binary TTSs with a narrow age spread and a spectral type distribution for the single stars identical to that of the primary stars in the binary sample. We used the strength of the 10\,\micron{} silicate emission feature derived from IRS/{\it Spitzer} spectra as a proxy for grain growth and the SED slope of circumstellar disks  as a proxy for dust settling. Our results can be summarized as follows: \\
\smallskip
-- there is no statistically significant difference between the distribution of 10\,\micron{} silicate emission features from single and binary systems. \\ 
-- the distribution of disk flaring is indistinguishable between the single and binary system samples. \\
\smallskip
These results show that stellar companions at projected separations of $\ga$\,10\,AU do not appreciably affect the degree of crystallinity nor the degree of grain growth. Based on the combination of these and other results we argue that the formation of planetesimals and possibly terrestrial planets is not inhibited in a circumstellar disk perturbed by a medium--separation stellar companion.



\acknowledgments
This work is based on observations made with 
the Spitzer Space Telescope, which is operated by the Jet Propulsion Laboratory, California 
Institute of Technology. We are pleased to acknowledge support through
NASA contract 1290778  and through the NASA Astrobiology Institute. We thank D. Padgett, PI of the {\it Spitzer} Legacy team mapping the Taurus Molecular Cloud,  for providing such a valuable dataset to the star and planet formation community. We thank the anonymous referee for thoughtful comments that helped to improve the manuscript.



{\it Facilities:} \facility{Spitzer Space Telescope}

\clearpage

\end{document}